\documentclass[preprint,pra,showpacs,nofootinbib]{revtex4}
\usepackage[mathscr]{euscript}
\usepackage[dvips]{pstcol}
\usepackage{graphicx}
\usepackage{float,epsfig}
\usepackage{dcolumn}
\usepackage{bm}
\usepackage{graphicx}
\usepackage{dcolumn}
\usepackage{amsmath,amssymb,amsthm}
\usepackage{subfigure}
\usepackage[colorlinks=true,linkcolor=blue]{hyperref}
\newcommand{\bea}{\begin{eqnarray}}
\newcommand{\eea}{\end{eqnarray}}
\newcommand{\beq}{\begin{equation}}
\newcommand{\eeq}{\end{equation}}

\def\/{\over}

\begin{document}

\title{Dynamical behavior and geometric phase for a circularly accelerated two-level atom}
\author{  Yao Jin,$^{1}$  Jiawei Hu$^{1}$, and Hongwei Yu$^{1,2,}$\footnote{Corresponding author} }
\affiliation{$^1$ Institute of Physics and Key Laboratory of Low
Dimensional Quantum Structures and Quantum
Control of Ministry of Education,\\
Hunan Normal University, Changsha, Hunan 410081, China \\
$^2$ Center for Nonlinear Science and Department of Physics, Ningbo
University, Ningbo, Zhejiang 315211, China}


\begin{abstract}
We study, in the framework of open quantum
systems, the time evolution of a circularly accelerated two-level atom
coupled  in
the multipolar scheme to a bath of fluctuating vacuum electromagnetic fields. We find that both the spontaneous transition rates and the geometric phase for a circularly accelerated atom do not exhibit a clear sign of thermal radiation characterized by the Planckian factor in contrast to the linear acceleration case. The spontaneous transition rates and effective temperature of the atom are examined in detail in the ultrarelativistic limit and are shown to be always larger than those in the linear acceleration case with the same proper acceleration. Unlike the effective temperature, the geometric phase is dependent on the initial atomic
states. We show that when the polar angle in Bloch sphere, $\theta$, that characterizes the initial state of the atom equals $\pi/{2}$,
the geometric phases acquired due to circular and linear acceleration are the same. However, for a generic state with an arbitrary $\theta$, the
phase will be in general different, and then we demonstrate in the ultrarelativistic limit  that the geometric phase acquired for the atom in circular motion is always larger than that in linear acceleration with same proper acceleration for $\theta\in(0,\frac{\pi}{2})\cup(\frac{\pi}{2},\pi)$.

\end{abstract}

\maketitle

\section{Introduction}

Unruh showed, using a model particle detector, that for a uniformly accelerated observer, the Minkowski vacuum is seen to be equivalent to a
thermal bath of Rindler particles at a temperature $T_U = a/2\pi$~\cite{Unruh}, where $a$ is the observer's proper acceleration. Since then, the Unruh effect has been extensively studied in various  contexts, such as proton decay~\cite{Matsas,Vanzella,Suzuki}, the bremsstrahlung effect associated with point charges~\cite{Higuchi92a,Higuchi92b}, quantum entanglement~\cite{entangle1,entangle2},  and the spontaneous excitation of accelerated atoms coupled to scalar~\cite{Audretsch94,Audretsch95,Audretsch95b}, electromagnetic~\cite{Passante97,Yu06} and Dirac fields~\cite{ZhouYu12}.
Recently, the geometric phase, which is first studied by Berry~\cite{Berry} for the adiabatic evolution of a closed system and then extended to open systems by others~\cite{Uhlmann,Sjoqvist,Singh,Tong,Wang}, was proposed to be utilized to detect the Unruh effect at lower accelerations first by Martin-Martinez et al.~\cite{Martin11} and then by us in a more realistic situation~\cite{Yu12}. At this point, let us note that the geometric phase of an open quantum system undergoing nonunitary evolution has recently been studied experimentally by measuring the decoherence factor of the off-diagonal elements of the reduced density matrix of the system~\cite{Cu10}.

Let us note that the Unruh effect is usually concerned with linearly accelerated observers. However, it is also interesting to study the case of observers in uniform circular motion, since the very large acceleration which is required to make the  Unruh effect experimentally observable is easier to achieve in circular motion. In 1980, Letaw and Pfautch investigated the quantization of scalar fields in rotating coordinates~\cite{Letaw80}. Soon afterward, the response of a circularly moving Unruh-DeWitt detector coupled to scalar fields was studied in Refs.~\cite{Letaw81} and \cite{Kim}. It was first pointed out by Bell and Leinaas in Ref. \cite{cir1} that circulating electrons in an external magnetic field can be utilized as two-level detectors to reveal the relation between acceleration and temperature. The population of the electron's energy state is modified by the  centripetal acceleration, which leads to the change of polarization of the electrons~\cite{cir1,cir3,cir4,cir5,cir6,cir7,cir8}. The effect of the fluctuations of the circulating electrons in vertical direction has been examined by further studies~\cite{cir5,cir8}. By using the formalism developed by Dalibard, Dupont-Roc and Cohen-Tannoudji(DDC), the contributions of vacuum fluctuations and radiation reaction to the spontaneous excitation of a circularly accelerated two-level system coupled to vacuum scalar fields in analogy with the electric dipole interaction has been studied in Ref.~\cite{Audretsch95b}.

In this paper, by treating a two-level atom as an open quantum system in a bath of fluctuating electromagnetic fields as opposed to scalar fields~\cite{Audretsch95b} in vacuum , we plan to study the time evolution of a circularly accelerated two-level system coupled to all vacuum modes of electromagnetic fields in a realistic multipolar coupling scheme~\cite{CPP95}. We  calculate the spontaneous transition rates and the geometric phase of the atom and compare the results with those for the linear acceleration~\cite{Yu12,Yu06} and the case of a thermal bath.

\section{the master equation }
The total Hamiltonian of the system consisting of a circularly accelerated two-level system coupled to fluctuating vacuum electromagnetic fields is given by
$
H=H_s+H_f+H'\;,
$
where $H_s$ is the Hamiltonian of
the atom, which takes the form $H_s={1\over
2}\,\hbar\omega_0\sigma_3$. Here $\sigma_3$ is the Pauli matrix
and $\omega_0$ is the energy level spacing of the atom. $H_f$ denotes the
Hamiltonian of the free electromagnetic field and the explicit expression
is not required here. The Hamiltonian that describes the interaction
between the atom and the  electromagnetic field in the multipolar
coupling scheme can be written as
\beq
 H'(\tau)=-e\textbf{r} \cdot
\textbf{E}(x(\tau))=-e\sum_{mn}\textbf{r}_{mn}\cdot
\textbf{E}(x(\tau))\sigma_{mn}\;,
\eeq
where {\it e} is the electron
electric charge, $e\,\bf r$  is the atomic electric dipole moment, and
${\bf E}(x)$ denotes the electric field strength.

We let $\rho_{tot}=\rho(0) \otimes |0\rangle\langle0|$  be the initial total density matrix of the system. Here $\rho(0)$ is the initial reduced density matrix of the atom, and $|0\rangle$ is the vacuum state of the field. In the frame of the atom, the evolution of the total density matrix $\rho_{tot}$ in the proper time $\tau$ reads
\begin{equation}
\frac{\partial\rho_{tot}(\tau)}{\partial\tau}=-{i\/\hbar}[H,\rho_{tot}(\tau)]\;.
\end{equation}
We assume that the interaction between the atom and  field is
weak. So, the evolution of the reduced
density matrix $\rho(\tau)$ can be written in the
Kossakowski-Lindblad form~\cite{Lindblad, pr5}
\begin{equation}\label{master}
{\partial\rho(\tau)\over \partial \tau}= -{i\/\hbar}\big[H_{\rm eff},\,
\rho(\tau)\big]
 + {\cal L}[\rho(\tau)]\ ,
\end{equation}
where
\begin{equation}
{\cal L}[\rho]={1\over2} \sum_{i,j=1}^3
a_{ij}\big[2\,\sigma_j\rho\,\sigma_i-\sigma_i\sigma_j\, \rho
-\rho\,\sigma_i\sigma_j\big]\ .
\end{equation}
The coefficients of the Kossakowski matrix $a_{ij}$ can be expressed as
\begin{equation}
a_{ij}=A\delta_{ij}-iB
\epsilon_{ijk}\delta_{k3}-A\delta_{i3}\delta_{j3}\;,
\end{equation}
with
\begin{equation}\label{abc}
A=\frac{1}{4}[{\cal {G}}(\omega_0)+{\cal{G}}(-\omega_0)]\;,\;~~
B=\frac{1}{4}[{\cal {G}}(\omega_0)-{\cal{G}}(-\omega_0)]\;.
\end{equation}
We introduce the two-point correlation function for electromagnetic fields as
\begin{equation}
G^{+}(x-y)={e^2\/\hbar^2} \sum_{i,j=1}^3\langle +|r_i|-\rangle\langle -|r_j|+\rangle\,\langle0|E_i(x)E_j(y)|0 \rangle\;.
\end{equation}
Here, $|+\rangle$, $|-\rangle$ denote the excited state and ground state of the atom respectively.  The Fourier and Hilbert transforms of the field correlation functions, ${\cal G}(\lambda)$ and ${\cal K}(\lambda)$, are defined as follows
\begin{equation}
{\cal G}(\lambda)=\int_{-\infty}^{\infty} d\Delta\tau \,
e^{i{\lambda}\Delta\tau}\, G^{+}\big(\Delta\tau\big)\; ,
\quad\quad
{\cal K}(\lambda)=\frac{P}{\pi
i}\int_{-\infty}^{\infty} d\omega\ \frac{ {\cal G}(\omega)
}{\omega-\lambda} \;.
\end{equation}
By absorbing the Lamb shift term, the effective Hamiltonian $H_{\rm eff}$ can written as
\begin{equation}\label{heff}
H_{\rm eff}=\frac{1}{2}\hbar\Omega\sigma_3={\hbar\over 2}\{\omega_0+{i\/2}[{\cal
K}(-\omega_0)-{\cal K}(\omega_0)]\}\,\sigma_3\;.
\end{equation}
Assuming that the initial state of the atom is
$|\psi(0)\rangle=\cos{\theta\/2}|+\rangle+\sin{\theta\/2}|-\rangle$, we obtain the time-dependent reduced density matrix:
\begin{equation}\label{dens}
\rho(\tau)=\left(
\begin{array}{ccc}
e^{-4A\tau}\cos^2{\theta\/2}+{B-A\/2A}(e^{-4A\tau}-1) & {1\/2}e^{-2A\tau-i\Omega\tau}\sin\theta\\ {1\/2}e^{-2A\tau+i\Omega\tau}\sin\theta & 1-e^{-4A\tau}\cos^2{\theta\/2}-{B-A\/2A}(e^{-4A\tau}-1)
\end{array}\right)\;.
\end{equation}

\section{transition rates and effective temperature of circularly accelerated atoms }

With the time-dependent reduced density matrix given, we can study the time evolution of atom observables. For an arbitrary Hermitian operator ${\cal O}$ that describes an observable of the atom, the evolution in time of its mean value can be expressed as
\beq
\langle {\cal O (\tau)}\rangle={\rm Tr}[{\cal O}\;\rho(\tau)]\;.\label{parameter}
\eeq
 If we let the Hermitian operator ${\cal O}$ be an admissible atomic state $\rho_f$, then Eq.~(\ref{parameter}) gives the transition probability ${\cal P}_{i\rightarrow f}$ from an initial atom state $\rho (0)\equiv\rho_i$ to the expected state $\rho_f$.
\beq
{\cal P}_{i\rightarrow f}={\rm Tr}[{\rho_f}\;\rho(\tau)]\;.
\eeq
If $\rho_i$ is the density matrix of the ground state with the polar angle in Bloch sphere $\theta=\pi$, and $\rho_f$ is the excited state $\theta=0$, then, with the help of Eq.~(\ref{dens}), we have
\beq
{\cal P}_{\uparrow}=\frac{A-B}{2A}(1-e^{-4A\tau})\;.
\eeq
As a result, the spontaneous excitation rate $\Gamma_\uparrow$, which corresponds to the transition probability per unit time at $\tau=0$ is
\beq
\Gamma_\uparrow=\frac{\partial}{\partial\tau}{\cal P}_{\uparrow}\bigg|_{\tau=0}=2A-2B={\cal G}(-\omega_0)\;.
\eeq
Similarly, we have
\beq
{\cal P}_{\downarrow}=\frac{A+B}{2A}(1-e^{-4A\tau})\;,
\eeq
and
\beq
\Gamma_\downarrow=\frac{\partial}{\partial\tau}{\cal P}_{\downarrow}\bigg|_{\tau=0}=2A+2B={\cal G}(\omega_0)\;
\eeq
for the spontaneous emission rate.

\begin{figure}[htp]
\begin{center}
\includegraphics{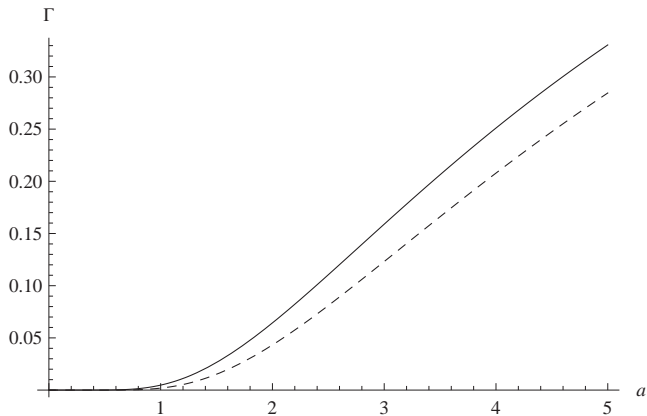}
\caption{Relative transition rate as a function of acceleration. Here
the acceleration is in the unit of transition frequency $\omega_0 c$. The solid and dashed
lines represent the circularly and linearly accelerated atom coupled to vacuum electromagnetic fields respectively. }
\label{fig2}
\end{center}
\end{figure}

Let us now calculate the spontaneous transition rates of a circularly accelerated two-level atom. 
The trajectory of the atom can be described as
\begin{eqnarray}
t(\tau)=\gamma\tau\ ,\ \ \ x(\tau)=R\cos {\gamma\tau v\/R}\ ,\ \
\ y(\tau)=R\sin {\gamma\tau v\/R}\ ,\ \ \ z(\tau)=0\label{tra}\;.\label{traj}
\end{eqnarray}
Here $R$ denotes the radius of the orbit, $v$ is the velocity of the atom, and $\gamma=1/\sqrt{1-{v^2/c^2}}$  is the usual Lorentz factor.  The centripetal acceleration in the frame of the atom is $a={\gamma^2v^2\/R}$. In order to obtain the transition rates, we need the field correlation functions, which can be found from the following two-point functions of the electric field
\begin{eqnarray}\label{vac-green}
\langle E_i(x(\tau))E_j(x(\tau'))\rangle={\hbar c\/4\pi^2\varepsilon_0}(\partial_0\partial_0^\prime\delta_{ij}-\partial_i\partial_j^\prime)
{1\/|{\bf x-x^\prime}|^2-(c\,t-c\,t'-i\varepsilon)^2}\;.
\end{eqnarray}
Applying the trajectory of the atom (\ref{traj}), one can easily obtain the field correlation function in the frame of the atom
\begin{eqnarray}\label{wightman1}
G^+(x,x')={e^2|\langle -|{\bf r}|+\rangle|^2\/\pi^2\varepsilon_0\hbar c^3} {1\/\{\gamma^2(\tau-\tau'-i\varepsilon)^2-({2v^2\gamma^2\/ac})^2\sin^2[{a\/2v\gamma}(\tau-\tau')]\}^2}\;,
\end{eqnarray}
which can be alternatively written as
\beq\label{green}
G^+(x,x')={e^2|\langle -|{\bf r}|+\rangle|^2\/\pi^2\varepsilon_0\hbar c^3} {1\/(\Delta\tau-i\varepsilon')^4[1+f(\Delta\tau)]^2}\;,
\eeq
with
\beq
f(\Delta\tau)={1\/12}\bigg[{a\/c}(\Delta\tau)\bigg]^2
-{c^2\/{360v^2\gamma^2}}\bigg[{a\/c}(\Delta\tau)\bigg]^4+\cdots\;.\label{f}
\eeq
Here  terms of all orders of $\Delta\tau$ are kept in Eq. (\ref{f}).
As is hard to find the explicit form of ${\cal G}(\omega_0)$ and ${\cal G}(-\omega_0)$, we now consider the ultrarelativistic limit $\gamma\gg1$~\cite{cir1}, in which
\begin{eqnarray}
G^+(x,x')={e^2|\langle -|{\bf r}|+\rangle|^2\/\pi^2\varepsilon_0\hbar c^3} {1\/(\tau-\tau'-i\varepsilon')^4\{1+{1\/12}[{a\/c}(\tau-\tau')]^2\}^2}\;.
\end{eqnarray}
Then, the Fourier transform of the field correlation function, which corresponds to the spontaneous emission rate, is given by
\begin{eqnarray}\label{fourier}
\Gamma_\downarrow={\cal G}(\omega_0)={\omega_0^3\,e^2|\langle -|{\bf r}|+\rangle|^2\/3\pi\varepsilon_0\hbar c^3} \bigg[1+\frac{a^2}{c^2\omega_0^2}+\bigg(\frac{a^2}{8c^2\omega_0^2}+\frac{5a^3}{16\sqrt{3}c^3\omega_0^3}\bigg)e^{-2\sqrt{3}\frac{\omega_0c}{a}}\bigg]\;.
\end{eqnarray}
Similarly, the spontaneous excitation rate is given by
\begin{eqnarray}\label{fourier2}
\Gamma_\uparrow={\cal G}(-\omega_0)={\omega_0^3\,e^2|\langle -|{\bf r}|+\rangle|^2\/3\pi\varepsilon_0\hbar c^3} \bigg(\frac{a^2}{8c^2\omega_0^2}+\frac{5a^3}{16\sqrt{3}c^3\omega_0^3}\bigg)e^{-2\sqrt{3}\frac{\omega_0c}{a}}\;.
\end{eqnarray}
So, unlike inertial atoms, circularly accelerated atoms in their ground state will be spontaneously excited. Unlike the linear acceleration case~\cite{Yu06}, the terms proportional to the Planckian factor are replaced by those proportional to an exponential term. This indicates that the radiation perceived by a circularly accelerated observer is not thermal. For $\frac{a}{c\omega_0}\ll 1$, though the leading part of the spontaneous decay rate in the present case and the linear acceleration case~\cite{Yu06} are same, the spontaneous excitation rate in these cases are different in this low acceleration limit. With the help of Eqs.~(\ref{fourier}) and Eq.~(\ref{fourier2}), we plot the relative transition rate $\Gamma={\Gamma_\uparrow}/{\Gamma_\downarrow}$ as a function of the  acceleration of the atom in Fig.~(\ref{fig2}) in comparison with the linear acceleration case~\cite{Yu06}. We find that the relative transition rate in the circular acceleration case is always larger than that in the linear acceleration case.

After evolving for a sufficiently long period of time $\tau\gg 1/(4A)$, the atom will be driven to a steady state, which is independent of the initial atomic state, and the density matrix (\ref{dens}) then becomes
\begin{equation}
\rho(\tau)=\frac{1}{{\cal{G}}(\omega_0)+{\cal{G}}(-\omega_0)}
\left(
\begin{array}{ccc}
{\cal{G}}(-\omega_{0}) & 0\\ 0 & {\cal{G}}(\omega_0)
\end{array}\right)\;.
\end{equation}
Let us note that an effective temperature for two-level atoms can be defined in terms of the relative transition rate as
\beq
k_B T_{\rm eff}=\hbar\omega_{0}[-\ln\Gamma(\omega_0)]^{-1}\;.
\eeq
Here $k_B$ denotes the Boltzmann constant. In the limit $\frac{a}{c\omega_0}\ll 1$, we have
\beq
\Gamma
\approx\frac{a^2}{8\omega_{0}^2c^2}e^{-2\sqrt{3}\frac{\omega_{0}c}{a}}\;,
\eeq
and the effective temperature can be simplified as $T_{\rm eff}=\frac{\hbar a}{2\sqrt{3}k_Bc}$
, which is higher by a factor $\frac{\pi}{\sqrt{3}}$ than the Unruh temperature for the linear acceleration.
\section{geometric phase of the circularly accelerated atom}

\begin{figure}[htbp]
\centering
\includegraphics{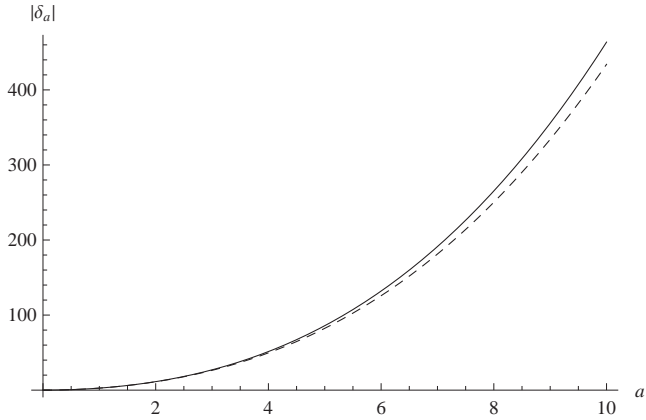}
\caption{Geometric phase as a function of acceleration for the initial atomic state with $\theta=\pi/4$. Here
the phases are in the units of $\frac{\pi^2\gamma_0}{4\omega_0}$
and the acceleration are in the units of $\omega_0c$. The solid and dashed
lines represent the circularly accelerated case and linear acceleration case respectively.}\label{fig1}
\end{figure}

The circularly accelerated atom which couples to fluctuating vacuum electromagnetic fields evolves nonunitarily and acquires a geometric phase during its evolution.  Here we calculate this geometric phase. For this purpose, let us note that the geometric phase for a mixed state under nonunitary
evolution is given by~\cite{Tong}
\beq\label{gp} \Phi_g=\arg
\left( \sum\limits_{k=1}^N \sqrt{\lambda_k(0)\lambda_k(T)}\langle
\phi_k(0)|\phi_k(T)\rangle e^{-\int_0^T \langle \phi_k(\tau)|\dot
\phi_k(\tau) \rangle d\tau} \right)\;,
\eeq
where $\lambda_k(\tau)$
and $|\phi_k(\tau)\rangle$ are the eigenvalues and eigenvectors of
the reduced density matrix $\rho(\tau)$. In order to get the
geometric phase, we first need to calculate the eigenvalues of the density
matrix (\ref{dens})
 \beq
\lambda_\pm(\tau)={1\/2}(1\pm\eta)\;,
 \eeq
 where $\eta=\sqrt{\rho_3^2+e^{-4A\tau}\sin^2\theta}$
and $\rho_3=e^{-4A\tau}\cos\theta+{B\over A}(e^{-4A\tau}-1)$.  Obviously, $\lambda_-(0)=0$. So, the contribution comes
only from the eigenvector corresponding to $\lambda_+$
\beq
|\phi_+(\tau)\rangle=\sin{\theta_{\tau}\/2}|+\rangle+\cos{\theta_{\tau}\/2}e^{i\Omega\tau}|-\rangle\;,
\eeq
where \beq\label{tan}
\tan{\theta_{\tau}\/2}=\sqrt{\eta+\rho_3\/\eta-\rho_3}\;. \eeq Then the
geometric phase can be calculated directly using Eq.~(\ref{gp}) as \bea\label{acc}
\Phi_g&=&-\Omega\int_0^T\cos^2{\theta_{\tau}\/2}\,d\tau\nonumber\\
&=&-\int_0^{T}{1\over2}\bigg(1-\frac{Q-Q\,e^{4A\tau}+\cos\theta}{\sqrt{e^{4A\tau}\sin^2\theta
+(Q-Q\,e^{4A\tau}+\cos\theta)^2}}\bigg)\,\Omega\,d\tau\;, \eea
where $Q=B/A$. It is clear that for $\theta=0$ and
$\theta=\pi$, which correspond to an initial excited state and a
ground state respectively, the geometric phase equals zero. We let $\gamma_0=e^2|\langle -|{\bf r}|+\rangle|^2\,\omega_0^3/3\pi\varepsilon_0\hbar c^3$, which is the
spontaneous emission rate of an inertial atom in the Minkowski vacuum. As is shown in Ref.~\cite{Yu12}, for small
$\gamma_0/\omega_0$, we can expand $\Phi_g$ for a single quasi-cycle to the first order as
\beq\label{gp-appr}
\Phi_g\approx -\pi(1-\cos\theta)-\frac{2\pi^2}{\omega_0}(2B+A\cos\theta)\sin^2\theta\;.
\eeq
When $\theta=\frac{\pi}{2}$, the initial state of the atom is $|\psi(0)\rangle=(|+\rangle+|-\rangle)/\sqrt2$, and the geometric phase can be simplified to
\beq
\Phi_g\approx -\pi-\frac{4\pi^2}{\omega_0}B\;,
\eeq
which  depends only on the parameter $B$.
Let us now calculate the geometric phase of a circularly accelerated two-level atom. 
In order to calculate the geometric phase, we need to compute the coefficients $A, B$   defined in Eq.~(\ref{abc}) which are determined by the Fourier
transforms of the field correlation function. It is interesting to note that
when $\theta=\pi/2$, the geometric phase depends only on $B=\frac{1}{4}[{\cal {G}}(\omega_0)-{\cal{G}}(-\omega_0)]$.
In Eq. (\ref{green}), the singularities resulting from $[1+f(\Delta\tau)]$ are symmetric with respect to the origin of the complex plane, so the residues at these points cancel in the calculation of the factor $B$, and the only contribution to $B$ comes from the residue of Eq. (\ref{green}) at the point $\Delta\tau=i\varepsilon'$ so that
\beq
B=2\pi i\;{\rm Res}\,[G^+(\Delta\tau)e^{i\omega_0\Delta\tau}]
|_{\Delta\tau=i\varepsilon'}
={1\/4}\gamma_0\,\bigg(1+{a^2\/c^2\omega_0^2}\bigg)\;,
\eeq
which is the same as that in the linear acceleration case~\cite{Yu12}. This means that for an initial state of the atom with $\theta=\pi/2$ the geometric phase for the circularly accelerated atom is the same as that for the linearly accelerated one with same acceleration. As for an arbitrary initial atomic state, the coefficient $A$ is also needed for the calculation of the geometric phase. However,
it is hard to find the explicit form of $A$, so we now consider the ultrarelativistic limit $\gamma\gg1$~\cite{cir1}. According to Eq.~(\ref{fourier}) and Eq.~(\ref{fourier2}), the coefficients of the Kossakowski matrix $a_{ij}$
can be written as
\bea
  A={1\/4}\gamma_0\,\bigg[1+\frac{a^2}{c^2\omega_0^2}+\bigg(\frac{a^2}{4c^2\omega_0^2}
  +\frac{5a^3}{8\sqrt{3}c^3\omega_0^3}\bigg)e^{-2\sqrt{3}\frac{\omega_0c}{a}}\bigg]\;,
  \quad\quad
  B={1\/4}\gamma_0\,\bigg(1+{a^2\/c^2\omega_0^2}\bigg)\;,\label{ABC}
\eea
and the effective level spacing of the atom as
\bea
\Omega=\omega_0&+&{\gamma_0P\/2\pi\omega_0^3}\int_0^\infty
d\omega\,\omega^3\bigg({1\/\omega+\omega_0}-{1\/\omega-\omega_0}\bigg)\nonumber\\
&&\quad\quad\times\bigg[1+\frac{a^2}{c^2\omega_0^2}+\bigg(\frac{a^2}{4c^2\omega_0^2}+\frac{5a^3}{8\sqrt{3}c^3\omega_0^3}\bigg)e^{-2\sqrt{3}\frac{\omega_0c}{a}}\bigg]\;.
\eea
Now the geometric phase can be obtained
after applying Eq.~(\ref{ABC}):
\bea\label{gp2}
\Phi_g^{(circular)}&\approx&-\pi(1-\cos\theta)
-\pi^2{\gamma_0\/2\omega_0}\sin^2\theta
\bigg[\bigg(1+{a^2\/c^2\omega_0^2}\bigg)(2+\cos\theta)\nonumber\\
&&+\bigg(\frac{a^2}{4c^2\omega_0^2}+\frac{5a^3}{8\sqrt{3}c^3\omega_0^3}\bigg)e^{-2\sqrt{3}\frac{\omega_0c}{a}}
\cos\theta\bigg]\;.
\eea
Here the first term $-\pi(1-\cos\theta)$ is the well known geometric phase for a closed system under unitary evolution, and the second term is a correction caused by the interaction between the accelerated atom and the environment. The correction to the geometric phase purely due to the circular motion can be found by subtracting the contribution of the inertial part $\Phi_g^{(inertial)}$ from Eq.~(\ref{gp2})
\bea\label{GPdiff} \delta^{(circular)}&=&\Phi_g^{(circular)}-\Phi_g^{(inertial)} \nonumber\\
&\approx&
-\pi^2{\gamma_0\/2\omega_0}\bigg[{a^2\/c^2\omega_0^2}(2+\cos\theta)
+
\bigg(\frac{a^2}{4c^2\omega_0^2}+\frac{5a^3}{8\sqrt{3}c^3\omega_0^3}\bigg)e^{-2\sqrt{3}\frac{\omega_0c}{a}}
\cos\theta \bigg]\sin^2\theta\;.
\eea
From the  result above, we can see that the geometric phase purely due to
the circular acceleration of the atom is determined by the acceleration
$a$ and the properties of the atom, which consist of the transition frequency
$\omega_0$, the spontaneous emission rate $\gamma_0$, and the initial
state of the atom characterized by $\theta$. Here the first term is same as that of the linear acceleration case~\cite{Yu12} and it becomes dominant  when $\frac{a}{c\omega_0}\ll 1$ for both the present case and the linear acceleration case. However, unlike the linear acceleration case, the second term is not proportional to the Planckian factor.  As a result, a
circularly accelerated atom does not feel a thermal radiation as a linearly accelerated one in terms of the geometric phase acquired. In other words, we do not see a clear sign of thermal radiation characterized by the Planckian factor in the expression of the geometric phase. As has already been pointed out, for atomic states with $\theta=0$ and
$\theta=\pi$, the geometric phase becomes zero, whereas
for $\theta=\pi/2$, both linear and circular acceleration lead to the same geometric phase acquired. Let us also note here that, for the case of a thermal bath~\cite{Yu12b}, the temperature induced phase correction is
\beq
\delta^{(thermal)}=-\pi^2{\gamma_0\/2\omega_0}\frac{2}{e^{\beta\omega_0}-1}\cos\theta\sin^2\theta\;,
\eeq
which vanishes for the initial state with $\theta=\pi/2$.
Therefore, the phase correction is significantly different for the accelerated cases and the thermal case when   $\theta=\pi/2$.
For a general initial atomic state with $\theta\in(0,\frac{\pi}{2})\cup(\frac{\pi}{2},\pi)$, the geometric phases acquired due to circular acceleration, linear acceleration  and  a thermal bath  differ from one another. For example, when we compare the present case with those of the linear acceleration and the thermal bath with $\beta=\frac{2\pi c}{a}$  for $\theta=\pi/4$, we can see that the phase acquired due to circular acceleration case is always larger than that due to  linear acceleration, as is shown graphically in Fig. (\ref{fig1}).
When $\frac{a}{c\omega_0}\gg1$,
the phase acquired for the three cases can be approximated as
\beq
\delta_a^{(circular)}\approx-\frac{5\sqrt{6}\pi^2\gamma_0}{192\omega_0}\frac{a^3}{\omega_0^3c^3}\;,\quad
\delta_a^{(linear)}\approx-\frac{\sqrt{2}\pi\gamma_0}{8\omega_0}\frac{a^3}{\omega_0^3c^3}\;,\quad
\delta^{(thermal)}\approx-\frac{\sqrt{2}\pi\gamma_0}{8\omega_0}\frac{a}{\omega_0c}\;.
\eeq
Therefore, for large acceleration, the leading terms of the phase  for the linear and circular acceleration cases are proportional to $\frac{a^3}{\omega_0^3c^3}$, which are much larger than that of the thermal case which is proportional to $\frac{a}{\omega_0c}$. In this limit, the geometric phase  in circular acceleration case is larger than that in the linear acceleration case by a factor $\frac{5\pi}{8\sqrt{3}}\approx1.1336$.

\section{conclusion}

In summary, we have calculated the spontaneous transition rates and the geometric phase of a circularly accelerated two-level atom which is in interaction with a bath of fluctuating quantum electromagnetic fields in vacuum. We find that both the spontaneous transition rates and the geometric phase for a circularly accelerated atom do not exhibit a clear sign of thermal radiation characterized by  the Planckian factor in contrast to the linear acceleration case~\cite{Yu12}. The relative spontaneous transition rates are calculated in the ultrarelativistic limit, and it has been found that the atom will be spontaneously excited and the relative transition rate  in the present case is always larger than that in the linear acceleration case~\cite{Yu06} with the same acceleration. The geometric phase is found to be dependent crucially on the initial state of the atom. If the initial state is the ground or excited state, then no geometric phase will be acquired by the atom no matter if it is in linear acceleration or circular motion; whereas,
if the initial state is a superposition of the ground and excited states with a polar angle $\theta$ in Bloch sphere, then when $\theta$ equals $\pi/2$, the geometric phases acquired in linear acceleration and in circular motion will be same. But, for a generic state of the atom with $\theta\in(0,\frac{\pi}{2})\cup(\frac{\pi}{2},\pi)$, we demonstrate in the ultrarelativistic limit  that the geometric phase acquired for the atom in circular motion is always larger than that in linear acceleration with the same proper acceleration.

\begin{acknowledgments}

This work was supported in part by the National Natural
Science Foundation of China under Grants No. 11075083,
No. 10935013, and No. 11375092; the Zhejiang Provincial
Natural Science Foundation of China under Grant
No. Z6100077; the National Basic Research Program of
China under Grant No. 2010CB832803; the Program for
Changjiang Scholars and Innovative Research Team in
University (PCSIRT, No. IRT0964); the Hunan Provincial
Natural Science Foundation of China under Grant
No. 11JJ7001 and  the SRFDP under Grant
No. 20124306110001.

\end{acknowledgments}


\end{document}